\documentclass{article}
\usepackage{amsmath}
\usepackage[psamsfonts]{amssymb}
\usepackage{cmmib57}
\usepackage{diagrams}
\newcommand{\bPf}{\par\vspace*{-4pt}\indent{\sc Proof.}\enskip}
\newcommand{\ePf}{\medskip}
\def\QED{\hskip0.1em\hfill\null\ \null\nobreak\hfill\kern3pt\vbox{\hrule\hbox
   {\vrule\kern1pt\vbox{\kern1.7pt\hbox{$\scriptscriptstyle{QED}$}
    \kern0.2pt}\kern1pt\vrule}\hrule}}

\def\END{\hskip0.1em\hfill\null\ \null\nobreak\hfill\kern3pt\vbox{\hrule\hbox
   {\vrule\kern1pt\vbox{\kern1.7pt\hbox{$\,\,\,\vspace{5pt}$}
    \kern0.2pt}\kern1pt\vrule}\hrule}}
\newtheorem{theorem}{Theorem}
\newtheorem{lemma}{Lemma}
\newtheorem{corollary}{Corollary}
\newtheorem{proposition}{Proposition}
\newtheorem{remark}{Remark}
\newtheorem{definition}{Definition}
\newtheorem{example}{Example}
\newcommand{\bCd}{\bEq\begin{CD}}
\newcommand{\eCd}{\end{CD}\eEq}
\newcommand{\bcd}{\beq\begin{CD}}
\newcommand{\ecd}{\end{CD}\eeq}
\newcommand{\ben}{\begin{enumerate}}
\newcommand{\een}{\end{enumerate}}
\newcommand{\bEq}{\begin{eqnarray}}
\newcommand{\eEq}{\end{eqnarray}}
\newcommand{\beq}{\begin{eqnarray*}}
\newcommand{\eeq}{\end{eqnarray*}}
\newcommand{\bDf}{\begin{definition}\em}
\newcommand{\eDf}{\end{definition}}
\newcommand{\bLm}{\begin{lemma}}
\newcommand{\eLm}{\end{lemma}}
\newcommand{\bPr}{\begin{proposition}}
\newcommand{\ePr}{\end{proposition}}
\newcommand{\bTh}{\begin{theorem}}
\newcommand{\eTh}{\end{theorem}}
\newcommand{\bCr}{\begin{corollary}}
\newcommand{\eCr}{\end{corollary}}
\newcommand{\bRm}{\begin{remark}\em}
\newcommand{\eRm}{\end{remark}}
\newcommand{\bEx}{\begin{example}\em}
\newcommand{\eEx}{\end{example}}

\newcommand{\ie}{{\em i.e$.$} }
\newcommand{\eg}{{\em e.g$.$} }
\newcommand{\R}{I\!\!R}

\newcommand{\mto}{\mapsto}

\newcommand{\der}{\partial}

\DeclareMathOperator{\im}{im}
\DeclareMathOperator{\byd}{{\raisebox{.1ex}{:}{=}}}

\newcommand{\ucar}[1]{\underset{#1}{\times}}

\newcommand{\owed}[1]{\overset{#1}{\wedge}}

\newcommand{\balp}{\boldsymbol{\alp}}
\newcommand{\bbet}{\boldsymbol{\bet}}
\newcommand{\bgam}{\boldsymbol{\gam}}

\newcommand{\blam}{\boldsymbol{\lam}}
\newcommand{\bmu}{\boldsymbol{\mu}}
\newcommand{\bnu}{\boldsymbol{\nu}}

\newcommand{\bsig}{\boldsymbol{\sig}}

\newcommand{\cA}{\mathcal{A}}

\newcommand{\cC}{\mathcal{C}}

\newcommand{\cE}{\mathcal{E}}

\newcommand{\cJ}{\mathcal{J}}

\newcommand{\cL}{\mathcal{L}}

\newcommand{\cT}{\mathcal{T}}

\newcommand{\cZ}{\mathcal{Z}}

\newcommand{\bx}{\boldsymbol{x}}
\newcommand{\by}{\boldsymbol{y}}

\newcommand{\bF}{\boldsymbol{F}}
\newcommand{\bG}{\boldsymbol{G}}

\newcommand{\bP}{\boldsymbol{P}}

\newcommand{\bU}{\boldsymbol{U}}

\newcommand{\bW}{\boldsymbol{W}}
\newcommand{\bX}{\boldsymbol{X}}
\newcommand{\bY}{\boldsymbol{Y}}

\newcommand{\sub}{\subset}

\newcommand{\wed}{\wedge}
\newcommand{\com}{\!\circ\!}
\newcommand{\ten}{\!\otimes\!}

\newcommand{\alp}{\alpha}
\newcommand{\bet}{\beta}
\newcommand{\gam}{\gamma}
\newcommand{\del}{\delta}
\newcommand{\eps}{\epsilon}
\newcommand{\zet}{\zeta}

\newcommand{\kap}{\kappa}
\newcommand{\lam}{\lambda}
\newcommand{\sig}{\sigma}

\newcommand{\ome}{\omega}
\newcommand{\Gam}{\Gamma}

\newcommand{\Lam}{\Lambda}

\newcommand{\vartht}{\vartheta}

\newcommand{\For}{{\Lambda}}
\newcommand{\Con}{{\mathcal{C}}}
\newcommand{\Hor}{{\mathcal{H}}}
\newcommand{\Var}{{\mathcal{V}}}
\newcommand{\Thd}{{\Theta}}
\title{\large{{\bf Global Generalized Bianchi Identities for Invariant Variational Problems on
Gauge-natural Bundles}}\thanks{
{\em Dedicated to Hartwig in occasion of his first birthday.}}}
\author{{\normalsize M.
Palese and E. Winterroth}\thanks{Both of them supported by
GNFM of INdAM and University of Torino.}
\\{\footnotesize Department of Mathematics,
University of Torino}
\\{\footnotesize Via C. Alberto 10, 10123 Torino, Italy}\\ 
{\footnotesize e--mails: 
{\sc palese@dm.unito.it, ekkehart@dm.unito.it}}}
\date{}
\begin{document}

\maketitle

\begin{abstract}
	
We derive both {\em local} and {\em global} generalized {\em Bianchi identities} 
for classical Lagrangian field theories on gauge-natural bundles. 
We show that globally defined generalized Bianchi identities
can be found without the {\em a priori}
introduction of a connection. The proof is based
on a {\em global} decomposition of the {\em variational Lie derivative} of 
the generalized Euler--Lagrange morphism
and the representation of the corresponding generalized Jacobi morphism on gauge-natural 
bundles. In particular, we show that {\em within} a gauge-natural invariant 
Lagrangian variational principle, the gauge-natural lift of 
infinitesimal principal automorphism {\em is not} intrinsically 
arbitrary.
As a consequence the existence of {\em canonical} global superpotentials 
for gauge-natural Noether conserved currents is
proved without resorting to additional structures.  
\medskip


\end{abstract}

\section{Introduction}

{\em Local generalized Bianchi identities} 
for geometric field theories were introduced 
\cite{AnBe51,Ber49,Ber58,Gol58,Kom59} 
 to get (after an 
integration by part procedure) a consistent
equation between {\em local} divergences within the first variation formula. It was also stressed that in the 
general theory of relativity these identities coincide with the 
contracted Bianchi identities for the curvature tensor of the 
pseudo-Riemannian metric. We recall that in the classical Lagrangian 
formulation of field theories the description of symmetries 
amounts to define suitable (vector) densities 
which generate the conserved currents; in all relevant physical theories
this densities are found to be the divergence of skew--symmetric 
(tensor) densities, 
which are called {\em superpotentials} for the conserved currents. 
It is also well known that the importance of superpotentials relies on 
the fact that they can be integrated to provide conserved quantities 
associated with the conserved currents {\em via} the Stokes theorem (see \eg \cite{FeFr91} and references quoted
therein).
 
Subsequently, many attempts to ``covariantize'' such a derivation of Bianchi 
identities and superpotentials 
have been made (see \eg 
\cite{AFFFR01,Cru85,FFR01,FeFr91,FFR03,JuSi00,Kat85} and the wide literature quoted therein) by resorting 
to background metrics or (fibered) connections 
used to perform covariant integration by parts to get covariant 
(variations of)
currents and superpotentials.
In particular, in \cite{FFP01} such a covariant derivation was implicitly assumed 
to hold true for
any choice of gauge-natural prolongations of principal connections equal 
to prolongations of
principal connections with respect to a linear symmetric 
connection on the basis manifold in the
sense of \cite{Kol75,Kol79,KolVir96}. 

In the present paper, we
 derive both {\em local} and {\em global} generalized Bianchi identities
for classical field theories by resorting to the gauge-natural 
invariance of the Lagrangian and {\em via} the application of the {\em Noether Theorems} \cite{Noe18}.  In particular we show
that invariant generalized Bianchi identities can be found without the {\em 
a priori} introduction of a connection. The
proof is based on a {\em global} decomposition of the {\em variational Lie derivative} of the generalized
Euler--Lagrange morphism involving the definition - and its representation - of a new morphism, 
the generalized gauge-natural Jacobi morphisms. It is in fact known that the second variation of a Lagrangian can be
formulated in terms of Lie derivative of the corresponding Euler--Lagrange morphism \cite{FrPa00,FrPa01,FPV02,GoSt73,Pa00}.  
As a consequence the existence of canonical, \ie {\em completely determined by the variational problem and its invariance
properties}, global superpotentials for gauge-natural Noether conserved currents is proved
without resorting to additional structures. 

Our general framework is the calculus of variations on finite order jet of fibered 
bundles. Fibered bundles will be assumed to be 
{\em gauge-natural bundles} (\ie jet prolongations of fiber bundles associated 
to some gauge-natural prolongation of a principal bundle $\bP$ 
\cite{Ec81,Ja90,Ja03,Kol75,Kol79,KMS93}) and 
variations of sections are (vertical) vector fields given by Lie 
derivatives of sections with respect to gauge-natural lifts of 
infinitesimal principal automorphisms (see \eg \cite{Ec81,JaMo92,KMS93}).

In this general geometric framework we 
shall in particular consider {\em finite order variational sequences
on gauge-natural bundles}.
The variational sequence on finite order jet prolongations of fibered manifolds was introduced by Krupka 
as the quotient
of the de Rham sequence of differential forms (defined on the prolongation of the fibered manifold) with respect to a
natural exact contact subsequence, chosen in such a way that the generalized Euler-Lagrange and Helmholtz-Sonin
mappings can be recognized as some of its quotient mappings \cite{Kru90,Kru93}. 
The representation of the quotient 
sheaves of the variational sequence as
sheaves of sections of tensor bundles
given in \cite{Vit98}
and previous results on {\em variational Lie derivatives} 
and Noether Theorems \cite{FFP01,FPV98a} will be used.
Furthermore, we relate the generalized Bianchi morphism to the second variation of the Lagrangian. A very fundamental
abstract result due to Kol\'a\v r concerning {\em  global} decomposition formulae of vertical morphisms, involved with 
the integration by parts procedure \cite{HoKo83,Kol83,Kol84}, will be 
a key tool. In 
order to apply this results, we stress linearity properties of 
the Lie derivative operator acting on sections of the gauge-natural bundle, which rely on properties 
of the gauge-natural lift of infinitesimal principal automorphisms. 
 The gauge-natural lift enables one  to define the generalized gauge-natural Jacobi morphism
(\ie a  generalized Jacobi morphism where the variation vector fields 
-- instead of general deformations -- are Lie derivatives 
of sections of the gauge-natural bundle with respect to gauge-natural lifts of infinitesimal 
automorphisms of the underlying principal bundle), the kernel of which plays a very fundamental role. 

The paper is structured as follows. In Section $2$ we state the geometric framework by defining the variational sequence
on gauge natural-bundles and by representing the Lie derivative of fibered morphisms on its quotient sheaves;
Section
$3$ is dedicated to the definition and the representation of the 
generalized gauge-natural Jacobi morphism associated with a generalized gauge-natural Lagrangian. 
We stress some linearity properties of
this morphism as a consequence of the properties of the gauge-natural lift of 
infinitesimal right-invariant automorphisms
of the underlying structure bundle. In Section $4$, 
by resorting to the Second Noether Theorem, we relate the generalized
Bianchi identities with the kernel of the gauge-natural Jacobi morphism. 
We prove that the generalized Bianchi identities hold true globally if and only if the vertical part of jet prolongations of
gauge-natural lifts of infinitesimal principal bundle automorphisms is 
in the kernel of the second variation, \ie of the generalized gauge-natural Jacobi morphism. 

Here, manifolds and maps between manifolds are $C^{\infty}$.
All morphisms of fibered manifolds (and hence bundles) will 
be morphisms over the identity of the base manifold, unless 
otherwise specified.

\section{Variational sequences on gauge-natural bundles}

\subsection{Jets of fibered manifolds}

In this Section we recall some basic facts about jet spaces.
We introduce jet spaces of a
fibered manifold and the sheaves of forms on the $s$--th order
jet space. Moreover, we recall the notion of horizontal and vertical
differential \cite{KMS93,MaMo83a,Sau89}.

Our framework is a fibered manifold $\pi : \bY \to \bX$,
with $\dim \bX = n$ and $\dim \bY = n+m$.

For $s \geq q \geq 0$ integers we are concerned with the $s$--jet space $J_s\bY$ of $s$--jet prolongations of (local) sections
of $\pi$; in particular, we set $J_0\bY \equiv \bY$. We recall the natural fiberings
$\pi^s_q: J_s\bY \to J_q\bY$, $s \geq q$, $\pi^s: J_s\bY \to \bX$, and,
among these, the {\em affine\/} fiberings $\pi^{s}_{s-1}$.
We denote with $V\bY$ the vector subbundle of the tangent
bundle $T\bY$ of vectors on $\bY$ which are vertical with respect
to the fibering $\pi$.

Charts on $\bY$ adapted to $\pi$ are denoted by $(x^\sig ,y^i)$.  Greek
indices $\sig ,\mu ,\dots$ run from $1$ to $n$ and they label basis
coordinates, while
Latin indices $i,j,\dots$ run from $1$ to $m$ and label fibre coordinates,
unless otherwise specified.  We denote by $(\der_\sig ,\der_i)$ and 
$(d^\sig, d^i)$ the local basis of vector fields and $1$--forms on $\bY$
induced by an adapted chart, respectively.
We denote multi--indices of dimension $n$ by boldface Greek letters such as
$\balp = (\alp_1, \dots, \alp_n)$, with $0 \leq \alp_\mu$,
$\mu=1,\ldots,n$; by an abuse
of notation, we denote with $\sig$ the multi--index such that
$\alp_{\mu}=0$, if $\mu\neq \sig$, $\alp_{\mu}= 1$, if
$\mu=\sig$.
We also set $|\balp| \byd \alp_{1} + \dots + \alp_{n}$ and $\balp ! \byd
\alp_{1}! \dots \alp_{n}!$.
The charts induced on $J_s\bY$ are denoted by $(x^\sig,y^i_{\balp})$, with $0
\leq |\balp| \leq s$; in particular, we set $y^i_{\bf{0}}
\equiv y^i$. The local vector fields and forms of $J_s\bY$ induced by
the above coordinates are denoted by $(\der^{\balp}_i)$ and $(d^i_{\balp})$,
respectively.

In the theory of variational sequences a fundamental role is played by the
{\em contact maps\/} on jet spaces (see \cite{Kru90,Kru93,MaMo83a,Vit98}).
Namely, for $s\geq 1$, we consider the natural complementary fibered
morphisms over $J_s\bY \to J_{s-1}\bY$
\beq
\mathcal{D} : J_s\bY \ucar{\bX} T\bX \to TJ_{s-1}\bY \,,
\qquad
\vartht : J_{s}\bY \ucar{J_{s-1}\bY} TJ_{s-1}\bY \to VJ_{s-1}\bY \,,
\eeq
with coordinate expressions, for $0 \leq |\balp| \leq s-1$, given by
\beq
\mathcal{D} &= d^\lam\ten {\mathcal{D}}_\lam = d^\lam\ten
(\der_\lam + y^j_{\balp+\lam}\der_j^{\balp}) \,,
\vartht &= \vartht^j_{\balp}\ten\der_j^{\balp} =
(d^j_{\balp}-y^j_{{\balp}+\lam}d^\lam)
\ten\der_j^{\balp} \,.
\eeq

The morphisms above induce the following natural splitting (and its dual):
\bEq
\label{jet connection}
J_{s}\bY\ucar{J_{s-1}\bY}T^*J_{s-1}\bY =\left(
J_s\bY\ucar{J_{s-1}\bY}T^*\bX\right) \oplus\cC^{*}_{s-1}[\bY]\,,
\eEq
where $\cC^{*}_{s-1}[\bY] \byd \im \vartht_s^*$ and
$\vartht_s^* : J_s\bY \ucar{J_{s-1}\bY} V^*J_{s-1}\bY \to
J_s\bY \ucar{J_{s-1}\bY} T^*J_{s-1}\bY \,$. We have the isomorphism 
$\cC^{*}_{s-1}[\bY] \simeq J_{s}\bY \ucar{J_{s-1}\bY} V^{*}J_{s-1}\bY$. The role of the splitting above will be fundamental in
the present paper.

If $f: J_{s}\bY \to \R$ is a function, then we set
$D_{\sig}f$ $\byd \mathcal{D}_{\sig} f$,
$D_{\balp+\sig}f$ $\byd D_{\sig} D_{\balp}f$, where $D_{\sig}$ is
the standard {\em formal derivative}.
Given a vector field $\Xi : J_{s}\bY \to TJ_{s}\bY$, the splitting
\eqref{jet connection} yields $\Xi \, \com \, \pi^{s+1}_{s} = \Xi_{H} + \Xi_{V}$
where, if $\Xi = \Xi^{\gam}\der_{\gam} + \Xi^i_{\balp}\der^{\balp}_i$, then we
have $\Xi_{H} = \Xi^{\gam}D_{\gam}$ and
$\Xi_{V} = (\Xi^i_{\balp} - y^i_{\balp + \gam}\Xi^{\gam}) 
\der^{\balp}_{i}$. We shall call $\Xi_{H}$ and $\Xi_{V}$ the 
horizontal and the vertical part of $\Xi$, respectively.

The splitting
\eqref{jet connection} induces also a decomposition of the
exterior differential on $\bY$,
$(\pi^{s}_{s-1})^*\com \,d = d_H + d_V$, where $d_H$ and $d_V$
are defined to be the {\em horizontal\/} and {\em vertical differential\/}.
The action of $d_H$ and $d_V$ on functions and $1$--forms
on $J_s\bY$ uniquely characterizes $d_H$ and $d_V$ (see, {\em e.g.},
\cite{Sau89,Vit98} for more details).
A {\em projectable vector field\/} on $\bY$ is defined to be a pair
$(\Xi,\xi)$, where $\Xi:\bY \to T\bY$ and $\xi: \bX \to T\bX$
are vector fields and $\Xi$ is a fibered morphism over $\xi$.
If there is no danger of confusion, we will denote simply by $\Xi$ a
projectable vector field $(\Xi,\xi)$.
A projectable vector field $(\Xi,\xi)$, with coordinate expression
$\Xi = \xi^{\sig}\der_{\sig} + \xi^i\der_{i}$,
$\xi = \xi^{\sig}\der_{\sig}$,
can be conveniently prolonged to a projectable vector field
$(j_{s}\Xi, \xi)$, whose coordinate expression turns out to be
\beq
j_{s}\Xi = \xi^{\sig}\der_{\sig} +
(D_{\balp}\xi^i - \sum_{\bbet + \bgam = \balp}
\frac{\balp !}{\bbet ! \bgam !} D_{\bbet}\xi^{\mu} \, y^i_{\bgam + \mu}
) \, \der_i^{\balp} \,,
\eeq
where $\bbet \neq 0$ and $0 \leq |\balp | \leq s$ (see \eg
\cite{Kru90,MaMo83a,Sau89,Vit98});
in particular, we have the following expressions
$(j_{s}\Xi)_{H} = \xi^{\sig} \, D_{\sig}$,
$(j_{s}\Xi)_{V} = D_{\balp}(\Xi_{V})^i \, \der_i^{\balp}$,
with $(\Xi_{V})^i = \xi^i - \, y^i_{\sig}\xi^{\sig}$, for the 
horizontal and the vertical part of $j_{s}\Xi$, respectively.
From now on, by an abuse of notation, we will write simply $j_{s}\Xi_{H}$ and
$j_{s}\Xi_{V}$. In particular, $j_{s}\Xi_{V}:J_{s+1}\bY\ucar{J_{s}\bY}J_{s}\bY\to J_{s+1}\bY\ucar{J_{s}\bY}J_{s}V\bY$.

We are interested in the case in which 
physical fields are assumed to be sections of a fibered bundle and the 
variations of sections are generated by suitable vector 
fields. More precisely, fibered bundles will be assumed to be 
{\em gauge-natural bundles} and 
variations of sections are (vertical) vector fields given by Lie 
derivatives of sections with respect to gauge-natural lifts of 
infinitesimal principal automorphisms.
Such geometric structures have been widely recognized to suitably describe 
so-called gauge-natural field theories, \ie physical theories in which
right-invariant infinitesimal automorphisms of the structure bundle $\bP$  
uniquely define the transformation laws of the fields themselves (see {\em e.g.} 
\cite{Ec81,KMS93}).
In the following, we shall develop a suitable geometrical setting which enables us to 
define and investigate the fundamental concept of 
conserved quantity in gauge-natural Lagrangian field theories.

\subsection{Gauge-natural prolongations}

First we shall recall some basic definitions and properties concerning
gauge-natural prolongations of (structure) principal bundles (for an extensive exposition see \eg \cite{KMS93} and
references therein; in the interesting paper \cite{Ja03} fundamental reduction theorems for
general linear connections on vector bundles are provided in the gauge-natural framework).

Let $\bP\to\bX$ be a principal bundle with structure group $\bG$.
Let $r\leq k$ be integers and $\bW^{(r,k)}\bP$ $\byd$ $J_{r}\bP\ucar{\bX}L_{k}(\bX)$, 
where $L_{k}(\bX)$ is the bundle of $k$--frames 
in $\bX$ \cite{Ec81,Ja90,KMS93}, $\bW^{(r,k)}\bG \byd J_{r}\bG\odot GL_{k}(n)$
the semidirect product with respect to the action of $GL_{k}(n)$ 
on $J_{r}\bG$ given by the 
jet composition and $GL_{k}(n)$ is the group of $k$--frames 
in $\R^{n}$. Here we denote by $J_{r}\bG$ the space of $(r,n)$-velocities on $\bG$.

Elements of $\bW^{(r,k)}\bP$ are given by $(j^{\bx}_{r}\gam,j^{0}_{k}t)$, 
with $\gam: \bX\to\bP$ a local section, $t:\R^{n}\to\bX$ locally invertible 
at zero, 
with $t(0) = \bx$, $\bx\in \bX$. Elements of $\bW^{(r,k)}\bG$ are
$(j^{0}_{r}g,j^{0}_{k}\alp)$, where
$g:\R^{n}\to\bG$, $\alp: \R^{n}\to\R^{n}$ locally 
invertible at zero, with $\alp(0) = 0$.  

\bRm
The bundle $\bW^{(r,k)}\bP$ is a principal bundle over $\bX$ with structure group
$\bW^{(r,k)}\bG$.
The right action of $\bW^{(r,k)}\bG$ on the fibers of $\bW^{(r,k)}\bP$
is defined by the composition of jets (see, {\em e.g.}, 
\cite{Ja90,KMS93}).
\END\eRm

\bDf
The principal bundle $\bW^{(r,k)}\bP$ (resp. the Lie group $\bW^{(r,k)}\bG$)
is said to be the {\em gauge-natural prolongation of order $(r,k)$ 
of $\bP$ (resp. of $\bG$)}.
\END\eDf

\bRm
Let $(\Phi,\phi)$ be a principal automorphism of $\bP$ \cite{KMS93}.
It can be prolonged in a
natural  way to a principal automorphism of
$\bW^{(r,k)}\bP$, defined by:
\beq
\bW^{(r,k)}(\Phi, \phi):(j_{r}^{\bx}\gam, j_{k}^{0}t) \mto 
(j^{\phi(\bx)}_{r}(\Phi \circ \gam \circ \phi^{-1}),j^{0}_{k}(\phi \circ 
t))\,.
\eeq

The induced automorphism $\bW^{(r,k)}(\Phi,\phi)$ is an equivariant 
automorphism of $\bW^{(r,k)}\bP$ with 
respect to the action of the structure group $\bW^{(r,k)}\bG$. 
We shall simply denote it by the same symbol $\Phi$,  
if there is no danger of confusion.
\END\eRm

\bDf
We define the {\em vector}
bundle over $\bX$ of right--invariant infinitesimal automorphisms of $\bP$
by setting $\cA = T\bP/\bG$. 

We also define the {\em vector} bundle  over $\bX$ of right invariant 
infinitesimal automorphisms of $\bW^{(r,k)}\bP$ by setting 
$\cA^{(r,k)} \byd T\bW^{(r,k)}\bP/\bW^{(r,k)}\bG$ ($r\leq k$).
\END\eDf

\bRm
We have the following projections
$\cA^{(r,k)} \to \cA^{(r',k')}$, $r\leq k$, $r'\leq k'$,
with $r\geq r'$, $s\geq s'$.\END
\eRm

\subsection{Gauge-natural bundles and lifts}

Let $\bF$ be any manifold and $\zet:\bW^{(r,k)}\bG\ucar{}\bF\to\bF$ be 
a left action of $\bW^{(r,k)}\bG$ on $\bF$. There is a naturally defined 
right action of $\bW^{(r,k)}\bG$ on $\bW^{(r,k)}\bP \times \bF$ so that
 we can associate in a standard way
to $\bW^{(r,k)}\bP$ the bundle, on the given basis $\bX$,
$\bY_{\zet} \byd \bW^{(r,k)}\bP\times_{\zet}\bF$.

\bDf
We say $(\bY_{\zet},\bX,\pi_{\zet};\bF,\bG)$ to be the 
{\em gauge-natural bundle} of order 
$(r,k)$ associated to the principal bundle $\bW^{(r,k)}\bP$ 
by means of the left action $\zet$ of the group 
$\bW^{(r,k)}\bG$ on the manifold $\bF$ \cite{Ec81,KMS93}. 
\END\eDf

\bRm
A principal automorphism $\Phi$ of $\bW^{(r,k)}\bP$ induces an 
automorphism of the gauge-natural bundle by:
\bEq
\Phi_{\zet}:\bY_{\zet}\to\bY_{\zet}: [(j^{x}_{r}\gam,j^{0}_{k}t), 
\hat{f}]_{\zet}\mto [\Phi(j^{x}_{r}\gam,j^{0}_{k}t), 
\hat{f}]_{\zet}\,, 
\eEq
where $\hat{f}\in \bF$ and $[\cdot, \cdot]_{\zet}$ is the equivalence class
induced by the action $\zet$.
\END\eRm

Denote by $\cT_{\bX}$ and $\cA^{(r,k)}$ the sheaf of
vector fields on $\bX$ and the sheaf of right invariant vector fields 
on $\bW^{(r,k)}\bP$, respectively. A functorial mapping $\mathfrak{G}$ is defined 
which lifts any right--invariant local automorphism $(\Phi,\phi)$ of the 
principal bundle $W^{(r,k)}\bP$ into a unique local automorphism 
$(\Phi_{\zet},\phi)$ of the associated bundle $\bY_{\zet}$. 
Its infinitesimal version associates to any $\bar{\Xi} \in \cA^{(r,k)}$,
projectable over $\xi \in \cT_{\bX}$, a unique {\em projectable} vector field 
$\hat{\Xi} \byd \mathfrak{G}(\bar{\Xi})$ on $\bY_{\zet}$ in the 
following way:
\bEq
\mathfrak{G} : \bY_{\zet} \ucar{\bX} \cA^{(r,k)} \to T\bY_{\zet} \,:
(\by,\bar{\Xi}) \mto \hat{\Xi} (\by) \,,
\eEq
where, for any $\by \in \bY_{\zet}$, one sets: $\hat{\Xi}(\by)=
\frac{d}{dt} [(\Phi_{\zet \,t})(\by)]_{t=0}$,
and $\Phi_{\zet \,t}$ denotes the (local) flow corresponding to the 
gauge-natural lift of $\Phi_{t}$.

This mapping fulfils the following properties:
\begin{enumerate}
\item $\mathfrak{G}$ is linear over $id_{\bY_{\zet}}$;
\item we have $T\pi_{\zet}\circ\mathfrak{G} = id_{T\bX}\circ 
\bar{\pi}^{(r,k)}$, 
where $\bar{\pi}^{(r,k)}$ is the natural projection
$\bY_{\zet}\ucar{\bX} 
\cA^{(r,k)} \to T\bX$;
\item for any pair $(\bar{\Lam},\bar{\Xi})$ of vector fields in 
$\cA^{(r,k)}$, we have
\beq
\mathfrak{G}([\bar{\Lam},\bar{\Xi}]) = [\mathfrak{G}(\bar{\Lam}), \mathfrak{G}(\bar{\Xi})]\,;
\eeq
\item we have the coordinate expression of $\mathfrak{G}$
\bEq\label{espressione}
\mathfrak{G} = d^\mu \ten \der_\mu + d^{A}_{\bnu}
\ten (\cZ^{i\bnu}_{A} \der_{i}) + d^{\nu}_{\blam}
\ten (\cZ^{i\blam}_{\nu} \der_{i}) \,,
\eEq
with $0<|\bnu|<k$, $1<|\blam|<r$ and 
$\cZ^{i\bnu}_{A}$, $\cZ^{i\blam}_{\nu}$ $\in C^{\infty}(\bY_{\zet})$ 
are suitable functions which depend on the bundle, precisely on the fibers (see \cite{KMS93}).
\end{enumerate}

\bDf
The map $\mathfrak{G}$ is called the {\em gauge-natural lifting 
functor}. 
The  projectable vector field $(\hat{\Xi},\xi)\equiv \mathfrak{G}((\bar{\Xi},\xi))$ is 
called the {\em 
gauge-natural lift} of $(\bar{\Xi},\xi)$ to the bundle $\bY_{\zet}$.\END
\eDf

\subsection{Lie derivative of sections of gauge-natural bundles}

Let $\gam$ be a (local) section of $\bY_{\zet}$, $\bar{\Xi}$ 
$\in \cA^{(r,k)}$ and $\hat\Xi$ its gauge-natural lift. 
Following \cite{KMS93} we
define a (local) section $\pounds_{\bar{\Xi}} \gam : \bX \to V\bY_{\zet}$, 
by setting:
$\pounds_{\bar{\Xi}} \gam = T\gam \circ \xi - \hat{\Xi} \circ \gam$. 

\bDf
The (local) section $\pounds_{\bar{\Xi}} \gam$ is called the {\em 
generalized Lie derivative} of $\gam$ along the vector field 
$\hat{\Xi}$.
\END\eDf

\bRm
This section is a vertical prolongation of $\gam$, \ie it satisfies
the property: $\nu_{\bY_{\zet}} \circ \pounds_{\Xi} \gam = \gam$, 
where $\nu_{\bY_{\zet}}$ is the projection 
$\nu_{\bY_{\zet}}: V\bY_{\zet} \to \bY_{\zet}$.
Its coordinate expression is given by
$(\pounds_{\bar{\Xi}}\gam)^{i} = \xi^{\sig} \der_{\sig} \gam ^{i} - 
\hat\Xi^{i}(\gam)$.
\eRm

\bRm\label{lie}
The Lie derivative operator acting on sections of gauge-natural 
bundles satisfies the following 
properties:
\begin{enumerate}\label{lie properties}
\item for any vector field $\bar{\Xi} \in \cA^{(r,k)}$, the 
mapping $\gam \mto \pounds_{\bar{\Xi}}\gam$ 
is a first--order quasilinear differential operator;
\item for any local section $\gam$ of $\bY_{\zet}$, the mapping 
$\bar{\Xi} \mto \pounds_{\bar{\Xi}}\gam$ 
is a linear differential operator;
\item by using the canonical 
isomorphism $VJ_{s}\bY_{\zet}\simeq J_{s}V\bY_{\zet}$, we have
$\pounds_{\bar{\Xi}}[j_{r}\gam] = j_{s} [\pounds_{\bar{\Xi}} \gam]$,
for any (local) section $\gam$ of $\bY_{\zet}$ and for any (local) 
vector field $\bar{\Xi}\in \cA^{(r,k)}$. 

We can regard $\pounds_{\bar{\Xi}}: J_{1}\bY_{\zet} \to V\bY_{\zet}$ 
as a morphism over the
basis $\bX$. In this case it is meaningful to consider the (standard) 
jet
prolongation of $\pounds_{\bar{\Xi}}$, denoted by 
$j_{s}\pounds_{\bar{\Xi}}: 
J_{s+1}\bY_{\zet} \to VJ_{s}\bY_{\zet}$.
Furthermore, we have $j_{s}\hat{\Xi}_{V}(\gam)=-\pounds_{j_{s}\bar{\Xi}}\gam$.
\item we can consider $\pounds$ as a bundle morphism:
\bEq
\pounds: J_{s+1}(\bY_{\zet} \ucar{\bX}
\cA^{(r,k)}) \to J_{s+1}\bY_{\zet}\ucar{J_{s}\bY_{\zet}}VJ_{s}\bY_{\zet}\,.
\eEq
\end{enumerate}
\END\eRm

\subsection{Variational sequences} 

For the sake of simplifying notation, sometimes, we will omit the subscript $\zet$, so 
that all our considerations shall refer to $\bY$ as a gauge-natural 
bundle as defined above.

We shall be here concerned with some distinguished sheaves of forms on jet
spaces \cite{Kru90,Kru93,Sau89,Vit98}.
Due to the topological triviality of the fibre of
$J_{s}\bY\to \bY$, we will consider sheaves on $J_{s}\bY$ with respect to
the topology generated by open sets of the kind
$\left( {\pi_0^s}\right)^{-1}(\bU)$, with $\bU\subset\bY$ open in $\bY$.

i. For $s \geq 0$, we consider the standard sheaves $\For^{p}_{s}$
of $p$--forms on $J_s\bY$.

ii. For $0 \leq q \leq s $, we consider the sheaves $\Hor^{p}_{(s,q)}$ and
$\Hor^{p}_{s}$ of {\em horizontal forms\/}, \ie of local {\em fibered morphisms} (following the well known
correspondence between forms and fibered morphisms over the basis manifold, see \eg \cite{KMS93}) over
$\pi^{s}_{q}$ and $\pi^{s}$ of the type
$\alp : J_s\bY \to \owed{p}T^*J_q\bY$ and $\bet : J_s\bY \to \owed{p}T^*\bX$,
respectively.

iii. For $0 \leq q < s$, we consider the subsheaf $\Con^{p}_{(s,q)}
\sub \Hor^{p}_{(s,q)}$ of {\em contact forms\/}, \ie
of sections $\alp \in \Hor^{p}_{(s,q)}$ with values into
$\owed{p} (\Con^{*}_{q}[\bY])$.
We have the distinguished subsheaf $\Con^{p}{_s} \sub
\Con^{p}_{(s+1,s)}$ of local fibered morphisms $\alp \in 
\Con^{p}_{(s+1,s)}$
such that $\alp = \owed{p}\vartht_{s+1}^* \,\com \,\Tilde{\alp}$, where
$\tilde{\alp}$ is a section of the fibration $J_{s+1}\bY \ucar{J_s\bY}$
$\owed{p}V^*J_s\bY$ $\to J_{s+1}\bY$ which projects down onto
$J_{s}\bY$.

\bRm
{\em Notice} that according to \cite{Kru90,Kru93,Vit98}, the fibered splitting
\eqref{jet connection} yields the {\em sheaf splitting}
$\Hor^{p}_{(s+1,s)}$ $=$ $\bigoplus_{t=0}^p$
$\Con^{p-t}_{(s+1,s)}$ $\wed\Hor^{t}_{s+1}$, which restricts to the inclusion
$\For^{p}_s$ $\sub$ $\bigoplus_{t=0}^{p}$
$\Con^{p-t}{_s}\wed\Hor^{t,}{_{s+1}^{h}}$,
where $\Hor^{p,}{_{s+1}^{h}}$ $\byd$ $h(\For^{p}_s)$ for $0 < p\leq n$ and the surjective map
$h$ is defined to be the restriction to $\For^{p}_{s}$ of the projection of
the above splitting onto the non--trivial summand with the highest
value of $t$.\END
\eRm

The induced sheaf splitting above plays here a fundamental role. We
stress again that, for any jet order $s$, it is induced by the 
natural contact structure on 
the affine bundle $\pi^{s+1}_{s}$. Since a variational problem 
(described by the corresponding action integral) 
is insensitive to the addition of any piece containing contact
factors, such an affine structure has been pointed out in \cite{Kru90} to be
fundamental for the description of the geometric structure of the Calculus of 
Variations on finite order jets of fibered
manifolds. This property reflects on the intrinsic structure of all objects 
defined and represented in the 
{\em variational sequence} of a given order which we are just going to introduce. 
In particular this holds true
for the generalized Jacobi morphism we will define and represent in Section $3$. 

\medskip

We shortly recall now the theory of variational sequences on finite order jet
spaces, as it was developed by D. Krupka in \cite{Kru90}.

By an abuse of notation, let us denote by $d\ker h$ the sheaf
generated by the presheaf $d\ker h$ in the standard way.
We set $\Thd^{*}_{s}$ $\byd$ $\ker h$ $+$
$d\ker h$.

In \cite{Kru90} 
it was proved that the following sequence is an exact resolution of the constant sheaf $\R_{\bY}$ over $\bY$:
\beq
\diagramstyle[size=1.3em]
\begin{diagram}
0 & \rTo & \R_{Y} & \rTo & \For^{0}_s & \rTo^{\cE_{0}} &
\For^{1}_s/\Thd^{1}_s & \rTo^{\cE_{1}} & \For^{2}_s/\Thd^{2}_s & \rTo^{\cE_{2}} &
\dots & \rTo^{\cE_{I-1}} & \For^{I}_s/\Thd^{I}_s & \rTo^{\cE_{I}} &
\For^{I+1}_s & \rTo^{d} & 0
\end{diagram}
\eeq

\bDf
The above sequence, where the highest integer $I$ depends on the dimension
of the fibers of $J_{s}\bY \to \bX$ (see, in particular, \cite{Kru90}), is said to be the $s$--th order
{\em variational sequence\/} associated with the fibered manifold
$\bY\to\bX$.
\END
\eDf

For practical purposes, specifically to deal with morphisms which have 
a well known interpretation within the Calculus of Variations, we 
shall limit ourselves to consider the truncated variational sequence: 
\beq
\diagramstyle[size=1.3em]
\begin{diagram}
0 &\rTo & \R_{Y} &\rTo & \Var^{0}_s & \rTo^{\cE_0} &
\Var^{1}_{s} & \rTo^{\cE_{1}} & \dots  & \rTo^{\cE_{n}} &
\Var^{n+1}_{s}  & \rTo^{\cE_{n+1}} & \cE_{n+1}(\Var^{n+1}_{s})  
& \rTo^{\cE_{n+2}} & 
0 \,,
\end{diagram}
\eeq
where, following \cite{Vit98}, the sheaves $\Var^{p}_{s}\byd 
\Con^{p-n}_{s}\wed\Hor^{n,}{_{s+1}^h}/h(d\ker h)$ with $0\leq p\leq n+2$ are 
suitable representations of the corresponding quotient 
sheaves in the variational sequence by means of sheaves of sections of vector 
bundles. 
We notice that in the following, to avoid confusion, sometimes (when 
the interpretation could be dubious) we shall denote with a 
subscript the relevant fibered bundle
on which the variational sequence is defined; \eg in the case above, we would write
$(\Var^{p}_{s})_{\bY}$.
 
Let $\alp\in\Con^{1}_s\wed\Hor^{n,}{_{s+1}^h} 
\sub \Var^{n+1}_{s+1}$. Then there is a unique pair of
sheaf morphisms (\cite{Kol83,KoVi03,Vit98})
\bEq\label{first variation}
E_{\alp} \in \Con^{1}_{(2s,0)}\wed\Hor^{n,}{_{2s+1}^{h}} \,,
\qquad
F_{\alp} \in \Con^{1}_{(2s,s)} \wed \Hor^{n,}{_{2s+1}^h} \,,
\eEq
such that 
$(\pi^{2s+1}_{s+1})^*\alp=E_{\alp}-F_{\alp}$,
and $F_\alp$ is {\em locally} of the form $F_{\alp} = d_{H}p_{\alp}$, with $p_{\alp}
\in \Con^{1}_{(2s-1,s-1)}\wed\Hor^{n-1}{_{2s}}$.

\bDf
Let $\gam \in \For^{n+1}_{s}$.
The morphism $E_{h(\gam)}\in\Var^{n+1}_{s}$ is called the 
{\em generalized Euler--Lagrange morphism} associated with 
$\gam$ and the operator $\cE_{n}$ is called the 
{\em generalized Euler--Lagrange operator}. 
Furthermore $p_{h(\gam)}$ is a generalized momentum associated with $E_{h(\gam)}$.
\END
\eDf

Let $\eta\in\Con^{1}_{s}\wed\Con^{1}_{(s,0)}\wed\Hor^{n,}{_{s+1}^{h}}\sub 
\Var^{n+2}_{s+1}$, 
then there is a unique morphism 
$$
K_{\eta} \in \Con^{1}_{(2s,s)}\otimes\Con^{1}_{(2s,0)}\wed\Hor^{n,}{_{2s+1}^{h}}
$$
such that, for all $\Xi:\bY\to V\bY$,
$
E_{{j_{s}\Xi}\rfloor \eta} = C^{1}_{1} (j_{2s}\Xi\ten K_{\eta})$,
where $C^1_1$ stands for tensor
contraction on the first factor and $\rfloor$ denotes inner product (see \cite{KoVi03,Vit98}). 
Furthermore, there is a unique pair of sheaf morphisms
\bEq\label{second}
H_{\eta} \in 
\Con^{1}_{(2s,s)}\wed\Con^{1}_{(2s,0)}\wed\Hor^{n,}{_{2s+1}^{h}} \,,
\quad
G_{\eta} \in \Con^{2}_{(2s,s)}\wed\Hor^{n,}{_{2s+1}^{h}} \,,
\eEq
such that 
${(\pi^{2s+1}_{s+1})}^*\eta = H_{\eta} - G_{\eta}$ and $H_{\eta} 
= \frac{1}{2} \, A(K_{\eta})$,
where $A$ stands for antisymmetrisation.
Moreover, $G_{\eta}$ is {\em locally} of the type $G_{\eta} = d_H q_{\eta}$, 
where 
$q_{\eta} \in \Con^{2}_{(2s-1,s-1)}\wed\Hor^{n-1}{_{2s}}$, hence 
$[\eta]=[H_{\eta}]$ \cite{KoVi03,Vit98}. 

\bDf
Let $\gam \in \For^{n+1}_{s}$. 
The morphism $H_{hd\gam}\equiv H_{[\cE_{n+1}(\gam)]}$, where square brackets denote equivalence class, is called the {\em
generalized Helmholtz  morphism} and the operator $\cE_{n+1}$ is called the {\em generalized Helmholtz operator}.
Furthermore $q_{hd\gam}\equiv q_{[\cE_{n+1}(\gam)]}$ is a generalized momentum
associated with the Helmholtz morphism.
\END
\eDf

\bRm
A section $\lam\in\Var^{n}_s$ is just a Lagrangian of order 
$(s+1)$ of 
the standard literature. 
Furthermore, 
$\cE_{n}(\lam) \in \Var^{n+1}_{s}$ coincides with the standard higher 
order Euler--Lagrange morphism associated with $\lam$.
\END
\eRm

\bRm
It is well known that it is always possible to find global morphisms $p_{h(\gam)}$ and
$q_{hd\gam}$ satisfying decomposition formulae above; however, this possibility depends in
general on the choice of a linear symmetric connection on the basis manifold (see
\cite{Alo98,Alo99,Kol93}). In the present paper, we shall avoid to perform such a choice {\em
a priori}, with the explicit intention of performing an invariant  derivation of generalized
Bianchi identities, which does not relay on an invariant decomposition involving (local)
divergences; that is in fact possible when resorting to the representation of the
Second Noether Theorem in the variational sequence, as shown by Theorem $3$ below. This is
due to the fact that in the quotient sheaves of the variational sequence contact forms and
horizontal differentials of contact forms of higher degree are factored out. It is also clear
that this will give, at least, prescriptions on the meaningful (within a fully gauge-natural
invariant variational problem) possible choices of connections to be used to derive
covariantly generalized Bianchi identities in the classical way.
\END
\eRm

\subsection{Variational Lie derivative}

In this Subsection, following essentially \cite{FPV98a}, we give a representation in the variational sequence of the standard
Lie derivative operator acting on fibered morphisms.
We consider a projectable vector field $(\Xi,\xi)$ on $\bY$
and take into account the Lie derivative operator $L_{{j}_{s}\Xi}$ with respect to the jet prolongation
$j_{s}\Xi$ of $\Xi$.
In fact, as well known, such a prolonged vector field preserves the fiberings $\pi_q^{s}$, 
$\pi^s$; hence it preserves the splitting
\eqref{jet connection}.
Thus we have
\beq
\cL_{{j}_{s}\Xi}:\Var^{p}_{s} \to
\Var^{p}_{s}:
[\alp] \mto \cL_{{j}_{s}\Xi}([\alp]) = [L_{{j}_{s}\Xi}\alp]\,.
\eeq

\bDf
Let $(\Xi,\xi)$ be a projectable vector field. We call the map
$\cL_{j_{s}\Xi}$ defined above the
{\em variational Lie derivative\/}.\END
\eDf

Variational Lie derivatives allow us to calculate infinitesimal
symmetries of forms in the variational sequence. In particular, we are
interested in
symmetries of generalized Lagrangians and Euler--Lagrange morphisms which will 
enable us to represent in this framework Noether Theorems as well as known results 
stated in the framework of geometric bundles (see \eg the fundamental papers 
by Trautman \cite{Tra62,Tra67,Tra96}).

\bRm
Let $s\leq q$. Then the inclusions $\For^{p}_{s}\sub\For^{p}_{q}$ and
$\Thd^{p}_{s}\sub\Thd^{p}_{q}$ yield the injective sheaf morphisms
(see \cite{Kru90})
$\chi^{q}_s :\left(\For^{p}_s/\Thd^{p}_s\right)\to
\left(\For^{p}_{q}/\Thd^{p}_{q}\right) :
[\alp ] \mto [{(\pi^{q}_s)}^*\alp ]$, hence the inclusions
$\kap^q_{s} : \Var^{p}_{s} \to \Var^{p}_{q}$ for $s \leq q$.
\END\eRm

The inclusions $\kap^q_{s}$ of the variational 
sequence of order $s$ in the variational sequence of order $q$
give rise to new representations of $\cL_{j_{s}\Xi}$ on $\Var^{p}_{q}$.
In particular, the following two results hold true \cite{FPV98a}.

\bTh\label{noether I} 
Let $[\alp] = h(\alp)$ $\in$ $\Var^{n}_{s}$. Then we
have {\em locally}
\beq
\kap^{2s+1}_{s} \com \cL_{j_{s}\Xi}(h(\alp)) =
\Xi_{V} \rfloor \cE_{n}(h(\alp))+
d_{H}(j_{2s}\Xi_{V} \rfloor p_{d_{V}h(\alp)}+ \xi \rfloor h(\alp))\,.
\eeq
\eTh

\bPf
 We have
\begin{align*}
\kap^{2s+1}_{s} \com \cL_{j_{s}\Xi}(h(\alp)) & =
h(L_{j_{s+1}\Xi}h(\alp)))
\\
& = d_{H}(j_{s+1}\Xi_{H} \rfloor h(\alp)) +
h(j_{s+2}\Xi_{V} \rfloor d_{V}h(\alp))
\\
& = d_{H}(\xi \rfloor h(\alp)) +
h(j_{2s+1}\Xi_{V} \rfloor (E_{d_{V}h(\alp)}+F_{d_{V}h(\alp)})) \,.
\end{align*}
Since $F_{d_{V}h(\alp)} = d_{H}p_{d_{V}h(\alp)}$ {\em locally}, then
\beq
\kap^{2s+1}_{s} \com \cL_{j_{s}\Xi}(h(\alp)) =
\Xi_{V} \rfloor \cE_{n}(h(\alp)) +
d_{H}(j_{2s}\Xi_{V} \rfloor p_{d_{V}h(\alp)}+
\xi \rfloor h(\alp)) \,.\QED
\eeq
\ePf

\bTh\label{GeneralJacobi} 
Let $\alp\in\For^{n+1}_{s}$. Then we have {\em globally}
\beq
\kap^{2s+1}_{s} \com \cL_{j_{s}\Xi} [\alp] =
\cE_{n}({j_{s+1}\Xi_{V} \rfloor h(\alp)}) +
C^{1}_{1}(j_{s}\Xi_{V}\ten K_{hd\alp}) \,.
\eeq
\eTh

\bPf
We have
\begin{align*}
\kap^{2s+1}_{s} \com \cL_{j_{s}\Xi} [\alp]
&= [\cE_{n}(j_{s+1}\Xi_{V} \rfloor h(\alp))+
j_{s+1}\Xi_{V} \rfloor d_{V}h(\alp)]
\\
&= \cE_{n}({j_{s+1}\Xi_{V} \rfloor h(\alp)}) +
C^{1}_{1}(j_{s}\Xi_{V}\ten K_{hd\alp}) \,.
\end{align*}
\QED
\ePf

\section{Variations and generalized Jacobi morphisms}

To proceed further, we now need to recall some previous results concerning 
the representation of {\em generalized Jacobi morphisms} in
variational sequences and their relation with the second variation of a 
generalized Lagrangian (\cite{FrPa00,FrPa01,FPV02,Pa00},
see also the fundamental paper \cite{GoSt73}).
In \cite{FrPa01} the relation between the classical variations 
and Lie derivatives of a 
Lagrangian with respect to (vertical)
variation vector fields was worked out and the {\em variational vertical derivative} was 
introduced as an operator acting on sections of sheaves of 
variational sequences, by showing that this operator is in fact a 
natural transformation, being a functor on the category of 
variational sequences.
We shall here introduce {\em formal variations} of a morphism as  
{\em multiparameter deformations} showing that this is equivalent to take iterated 
variational Lie derivatives with respect to (vertical) variation vector fields. 
Our aim is to relate, on the basis of relations provided 
by Corollary \ref{kakka} and Proposition \ref{x} below,
 the second variation of 
the Lagrangian $\lam$ to the
Lie derivative of the associated Euler--Lagrange morphism and to the generalized
Bianchi morphism, defined by Eq. \eqref{key} in Subsection \ref{key1} below.
\medskip

We recall (see \cite{FrPa00,FrPa01,FPV02}) a Lemma which relates the $i$--th variation with the iterated 
Lie derivative of the morphism itself. 
Furthermore, following \cite{Kol83}, we recall the relation between the 
variation of the morphism and the vertical exterior differential.

\bDf\label{var}
Let $\alp:J_{s}\bY\to
\owed{p}T^*J_{s}\bY$. Let $\psi^{k}_{t_{k}}$, with $1\leq k\leq i$, be the 
flows generated by an $i$--tuple
$(\Xi_{1},\ldots,\Xi_{i})$ of (vertical, although actually it is enough that 
they are projectable) vector fields on $\bY$ and let $\Gam_{i}$ be the $i$--th
{\em formal variation} generated by the $\Xi_{k}$'s (to which we shall 
refer as variation vector fields)  and defined, 
for each $\by\in\bY$, by $\Gam_{i}(t_{1},\ldots,t_{i})(\by)= 
\psi^{i}_{t_{i}}\circ \ldots \circ \psi^{1}_{t_{1}}(\by)$. We define the $i$--th formal variation of the morphism $\alp$ to
be
\bEq
\del^{i}\alp \byd \frac{\der^{i}}{\der t_{1}\ldots 
\der t_{i}}\big |_{t_{1},\ldots,
t_{i}=0}(\alp\circ j_{s}\Gam_{i}(t_{1},\ldots,t_{i})(\by)) \,.\END
\eEq
\eDf

The following two Lemmas state the relation between the $i$--th formal variation 
of a morphism and its iterated Lie derivative 
\cite{FrPa00,FrPa01,FPV02,GoSt73}.

\bLm\label{LIE}
Let $\alp: J_{s}\bY\to
\owed{p}T^*J_{s}\bY$ and $L_{j_{s}\Xi_{k}}$ be the Lie derivative 
operator acting on differential fibered morphism.

Let 
$\Gam_{i}$ be the $i$--th formal variation generated by variation 
vector fields $\Xi_{k}$, $1\leq k\leq i$ on $\bY$. Then we have
\bEq
\del^{i} \alp = L_{j_{s}\Xi_{1}} \ldots L_{j_{s}\Xi_{i}} \alp \,.
\eEq
\eLm

\bLm\label{xx}
Let $\Xi$ be a variation vector field on $\bY$
and
$\lam\in\For^{n}_{s}$. Then we have 
$\del\lam = j_{s}\Xi\rfloor d_{V}\lam$ \cite{Kol83}.
\eLm

\bRm\label{iterate}
Owing to the linearity properties of $d_{V}\lam$, we can think of the operator $\del$ as a
{\em linear} morphism with respect to the vector bundle structure 
$J_{s}V\bY\to\bX$, so that we can write $\del\alp: J_{s}\bY\to J_{s}V^{*}\bY \wed \owed{n}T^*\bX$.
This property can be obviously iterated for each integer $i$, so that one can analogously 
define an
$i$--linear morphism $\del^{i}$. In particular, we have $\del^{2}\alp: J_{s}\bY\ucar{\bX}J_{s}V\bY\to
J_{s}V^{*}\bY\ten J_{s}V^{*}\bY\wed
\owed{n}T^*\bX$. \END
\eRm

For notational convenience and by an abuse of notation, in the sequel we shall denote with the same symbol
an object defined on the vertical prolongation $V\bY$ as well as the corresponding one defined on the iterated vertical
prolongation $V(V\bY)$, whenever there is no danger of confusion. 

\bCr\label{kakka}
Let $\lam\in (\For^{n}_{s})_{\bY}$. 
Let $\Xi_{1}$, $\Xi_{2}$ be two variation vector fields on
$\bY$ generating the formal variation $\Gam_{2}$. Then we have
\bEq
\del^{2}\lam &=&
j_{2s}\Xi_{2}\rfloor\cE_{n}(\del\lam)+d_{H}(j_{2s}\Xi_{2}\rfloor
p_{d_{V}\del\lam})\label{functor}
\\
&=& \del (j_{2s}\Xi_{1}\rfloor \cE_{n}(\lam)+d_{H}(j_{2s}\Xi_{1}\rfloor
p_{d_{V}\lam}))
\\
&=& j_{2s}\Xi_{2}\rfloor\cE_{n}((j_{2s}\Xi_{1}\rfloor\cE_{n}(\lam))+d_{H}(j_{2s+1}\Xi_{2}\rfloor
p_{d_{V}(\Xi_{1}\rfloor\cE_{n}(\lam))}+ \\
&+& \del(j_{2s}\Xi_{1}\rfloor
p_{d_{V}\lam}))
\,.
\eEq
\eCr

\bPf
We apply Lemma \ref{xx} and decomposition provided by Theorem 
\ref{noether I}. Furthermore, $d_{H}\del=\del d_{H}$, 
which follows directly from the analogous naturality property of the 
Lie derivative operator.
\ePf\QED

\bRm\label{boh}
From the relations above we also infer, of course, that
\beq
& &j_{2s}\Xi_{2}\rfloor\cE_{n}((j_{2s}\Xi_{1}
\rfloor\cE_{n}(\lam))+d_{H}(j_{2s+1}\Xi_{2}\rfloor
p_{d_{V}(\Xi_{1}\rfloor\cE_{n}(\lam))})= \\
& & =\del (j_{2s}\Xi_{1}\rfloor 
\cE_{n}(\lam))= 
\del^{2}\lam \,,\\
& &j_{2s}\Xi_{2}\rfloor\cE_{n}((j_{2s}\Xi_{1}\rfloor\cE_{n}(\lam))+
d_{H}(\del(j_{2s}\Xi_{1}\rfloor
p_{d_{V}\lam}))=j_{2s}\Xi_{2}\rfloor\cE_{n}(\del\lam) \,.\END
\eeq
\eRm
\subsection{Variational vertical derivatives and generalized Jacobi morphisms}

In this Section we restrict our attention to morphisms which are (identified with) 
sections of sheaves in the
variational sequence. 
We shall recall some results of ours \cite{FrPa00,FrPa01} by defining 
the $i$-th {\em variational vertical derivative} 
of morphisms.

Let $\alp\in (\Var^{n}_{s})_{\bY}$. We have
\beq
\del^{i}[\alp]\byd [\del^{i}\alp] =[L_{\Xi_{i}} \ldots
L_{\Xi_{1}}\alp]=\cL_{\Xi_{i}} \ldots 
\cL_{\Xi_{1}}[\alp]\,.
\eeq

\bDf
We call the operator $\del^{i}$ the {\em $i$--th variational vertical 
derivative}. \END
\eDf

In \cite{FrPa01} the variational vertical derivative was 
introduced as an operator acting on sections of sheaves of 
variational sequences, by showing that this operator is in fact a 
natural transformation, being a functor on the category of 
variational sequences as it can be summarized by the following commutative diagram.

\beq
\diagramstyle[size=2.3em]
\begin{diagram}
 \dots & \rTo^{d_{H}} &
(\Var^{n}_s)_{\bY} & \rTo^{\cE_{n}} & (\Var^{n+1}_s)_{\bY} & \rTo^{\cE_{n+1}} &
\cE_{n+1}(\Var^{n+1}_s)_{\bY} & \rTo^{\cE_{n+2}} & 0
\\
 && \dTo^{\del} && \dTo^{\del} && \dTo^{\del} && 
\\
 \dots & \rTo^{d_{H}} &
(\Var^{n}_s)_{\bY\ucar{\bX}V\bY} & \rTo^{\cE_{n}} & (\Var^{n+1}_s)_{\bY\ucar{\bX}V\bY} & \rTo^{\cE_{n+1}} &
\cE_{n+1}((\Var^{n+1}_s)_{\bY\ucar{\bX}V\bY}) & \rTo^{\cE_{n+2}} & 0
\\
 && \dTo^{\del} && \dTo^{\del} && \dTo^{\del} && 
\\
 \dots & \rTo^{d_{H}} &
(\Var^{n}_s)_{\bY\ucar{\bX}V(V\bY)} & \rTo^{\cE_{n}} & (\Var^{n+1}_s)_{\bY\ucar{\bX}V(V\bY)} &
\rTo^{\cE_{n+1}} & \cE_{n+1}((\Var^{n+1}_s)_{\bY\ucar{\bX}V(V\bY)}) & 
\rTo^{\cE_{n+2}} & 0
\\
 && \dTo^{\del} && \dTo^{\del} && \dTo^{\del} && 
\\
 \dots & \dots &
\dots & \dots & \dots & \dots &
\dots & \dots & \dots
\end{diagram}
\eeq

As a straightforward consequence we have the following
characterization of the second variation of a generalized Lagrangian 
in the variational sequence.

\bPr\label{x}
Let $\lam\in (\Var^{n}_{s})_{\bY}$ and let $\Xi$ be a variation vector 
field; then we have
\bEq
\del^{2}\lam = [\cE_{n}(j_{2s}\Xi \rfloor h\del\lam)
+C^{1}_{1} (j_{2s}\Xi \ten K_{hd\del\lam})] \,.
\eEq
\ePr

\bPf
Since $\del\lam\in (\Var^{n}_{s})_{V\bY}$, by Remark \ref{iterate} we have that 
$\del\lam\in (\Var^{n+1}_{s})_{\bY}$ and
then $hd\del\lam \in\cE_{n+1}(\Var^{n+1}_{s})_{\bY} \sub (\Var^{n+2}_{s})_{\bY}$; 
thus the assertion follows by a straightforward application of 
Theorem \ref{GeneralJacobi}. Notice that here $C^{1}_{1} (j_{2s}\Xi\ten K_{hd\del\lam}): 
J_{2s}\bY\ucar{\bX}J_{2s}V\bY \to V^{*}\bY\wed \owed{n}T^*\bX$. 
\ePf\QED

\bRm\label{Fund}
Let $l\geq 0$ and let $\bF$ be any vector bundle over $\bX$. 
Let $\alp: J_{l}(\bY \ucar{\bX} \bF)$ $\to$
$\owed{p}T^{*}\bX$  be a linear morphism with respect to the 
fibering $J_{l}\bY \ucar{\bX} J_{l}\bF$ $\to$ 
$J_{l}\bY$ and let ${\hat{D}}_{H}$ be the horizontal differential 
on $\bY \ucar{\bX} \bF$. We can uniquely write $\alp$ as
\beq
\overline{\alp}\equiv \alp: J_{l}\bY \to 
\cC^{*}_{l}[\bF]\wed (\owed{p}T^{*}\bX)\,.
\eeq
Then $\overline{{\hat{D}}_{H}\alp} = {\hat{D}}_{H}\overline{\alp}$ (this 
property was pointed out in  
\cite{FPV98b}).\END
\eRm

\bLm Let $\Xi$ be a variation vector field. 
Let $\chi(\lam,\Xi) \byd C^{1}_{1} (j_{2s}\Xi\ten K_{hd\del\lam}) \equiv 
E_{j_{s}\Xi\rfloor h(d\del\lam)}$ and let 
$\tilde{D}_{H}$ be the horizontal differential on 
$\bY\ucar{\bX} V\bY$. 
We can see $\chi(\lam,\Xi)$ as an extended morphism 
$\chi(\lam,\Xi): J_{2s}(\bY\ucar{\bX} V\bY)
\to J_{2s}V^{*}(\bY\ucar{\bX} V\bY)\ten V^{*}\bY\wed (\owed{n}T^{*}\bX)$
satisfying $\tilde{D}_{H}\chi(\lam,\Xi)=0$.
\eLm

\bPf
The morphism $\chi(\lam,\Xi): J_{2s}(\bY\ucar{\bX} V\bY)
\to V^{*}\bY\wed (\owed{n}T^{*}\bX)$ is a linear morphism with respect 
to the projection $J_{2s}(\bY\ucar{\bX} V\bY)\to J_{2s}\bY$ (see 
Remark \ref{iterate}), then we can apply the Remark above, so that $\chi(\lam,\Xi): J_{2s}\bY
\to J_{2s}V^{*}(V\bY)\ten V^{*}\bY\wed (\owed{n}T^{*}\bX)\simeq 
J_{2s}V^{*}\bY\ten J_{2s}V^{*}\bY\ten V^{*}\bY\wed (\owed{n}T^{*}\bX)$ 
and again by linearity we get $\chi(\lam,\Xi): J_{2s}(\bY\ucar{\bX} V\bY)
\to J_{2s}V^{*}\bY\ten V^{*}\bY\wed (\owed{n}T^{*}\bX)$. \QED
\ePf

The following Lemma is an application of an abstract result, due to Hor\'ak and Kol\'a\v r
\cite{HoKo83,Kol83,Kol84}, concerning a global decomposition formula for vertical morphisms. 
\bLm\label{kolar} 
Let $\Xi$ be a variation vector field. \\
Let 
$\chi(\lam,\Xi)$ as in the above Lemma.
Then we have 
$(\pi^{4s+1}_{2s+1})^{*}\chi(\lam,\Xi) = E_{\chi(\lam,\Xi)} + F_{\chi(\lam,\Xi)}$, where
\bEq
E_{\chi(\lam,\Xi)}: 
J_{4s}(\bY\ucar{\bX} V\bY)
\to \Con^{*}_{0}[\bY]\ten \Con^{*}_{0}[\bY]\wed \owed{n}T^{*}\bX\,,
\eEq
and {\em locally}, $F_{\chi(\lam,\Xi)} = \tilde{D}_{H}M_{\chi(\lam,\Xi)}$, with
\beq M_{\chi(\lam,\Xi)}: J_{4s-1}(\bY\ucar{\bX} V\bY) \to
\Con^{*}_{2s-1}[\bY]\ten \Con^{*}_{0}[\bY] \wed \owed{n-1}T^{*}\bX\,.
\eeq
\eLm

\bPf
Following {\em e.g.} \cite{Kol83,Kol84,Vit98}, the {\em global} morphisms 
$E_{\chi(\lam,\Xi)}$ and $F_{\chi(\lam,\Xi)}$
can be evaluated by means of a backwards procedure.
\QED
\ePf

\bDf
We call the morphism $\cJ (\lam,\Xi) \byd E_{\chi(\lam,\Xi)}$ 
the {\em generalized Jacobi morphism}  associated with the
Lagrangian
$\lam$.
\END\eDf
 
\bEx
Let us write explicitly the coordinate expression of
$\cJ (\lam,\Xi)$. By functoriality of $\del$, we have
$h(d\del\lam)=h(\del d\lam)$. Let now locally $\lam=L\ome$, where $L$ is a function of 
$J_{s}\bY$ and $\ome$ 
a volume form on $\bX$,
then
$d\lam=\der^{\balp}_{i}(L)d^{i}_{\balp}\wed\ome$ and 
$\del d\lam=\der^{\bsig}_{j}(\der^{\balp}_{i} L)d^{j}_{\bsig}\ten
d^{i}_{\balp}\wed \ome$,
thus finally $h(\del d\lam)$
$=$ $h(d\del \lam)$ $=$ $\der_{j}(\der^{\balp}_{i} L)d^{j}_{\balp}\ten d^{i}
\wed \ome$.
As a consequence we have, with $0\leq |\bmu|, |\balp|, |\bsig|\leq 2s+1$:
\begin{eqnarray*}
& &\chi(\lam,\Xi) = \\
& & = D_{\bsig}
\Xi^{l}_{V}\left(\der_{j}(\der^{\bmu}_{i}L) - \sum_{|\balp | 
= 0}^{s-|\bmu |}
(-1)^{|\bmu +\balp |} \frac{(\bmu + 
\balp)!}{\bmu ! \balp !} 
D_{\balp}\der^{\balp}_{j}(\der^{\bmu}_{i}L)\right) \der^{\bsig}_{l} \ten
\vartht^j_{\bmu}\ten\vartht^i \wed \ome \doteq \\
& &\doteq
\chi^{l \bmu }_{\bsig j i } \der^{\bsig}_{l} \ten \vartht^j_{\bmu}\ten\vartht^i \wed \ome \,;
\end{eqnarray*}
and by the Lemma above, we
get (up to divergencies):
\bEq\label{JJJ}
 \cJ(\lam,\Xi)=(-1)^{|\balp|}D_{\balp}\,
\chi^{l \balp}_{ \bsig j i} \der^{\bsig}_{l} \ten \vartht^j\ten \vartht^i\wed
\ome \,.\END
\eEq
\eEx

\subsection{Generalized {\em gauge-natural} Jacobi morphisms}

We intend now to specify the just mentioned 
results and definitions concerning the Jacobi morphism 
by considering as variation vector fields the vertical parts of prolongations of
gauge-natural lifts of infinitesimal principal automorphisms to the gauge-natural bundle $\bY_{\zet}$.
Owing to linearity properties of 
the Lie derivative of sections and taking into account the fact that, as we already recalled,
$j_{s}\hat{\Xi}_{V}=-\pounds_{j_{s}\bar{\Xi}}$, we can state the following important results.

\medskip

Recall (see \cite{KMS93}, Proposition 15.5) that the jet prolongation of order $s$ of $\bY_{\zet}$ is a gauge-natural
bundle itself associated to some principal prolongation of order $(r+s,k+s)$ of the underling principal bundle $\bP$. 
Let $\bar{\Xi}\in \cA^{(r,k)}$ and $\hat{\Xi}\byd\mathfrak{G}(\bar{\Xi})$ the 
corresponding gauge-natural lift to
$\bY_{\zet}$. Let $j_{s}\hat{\Xi}$ be the $s$-jet prolongation of $\hat{\Xi}$ 
which is a vector field on $J_{s}\bY_{\zet}$. It turns out then that it
is a gauge natural lift of $\bar{\Xi}$ too, \ie 
$j_{s}\mathfrak{G}(\bar{\Xi})=\mathfrak{G}(j_{s}\bar{\Xi})$. Let us consider $j_{s}{\hat{\Xi}}_{V}$, \ie the vertical part according to the
splitting \eqref{jet connection}. We shall denote by $j_{s}\bar{\Xi}_{V}$ the induced section of the vector bundle
$\cA^{(r+s,k+s)}$. The set of all sections of this kind defines a vector subbundle of $J_{s}\cA^{(r,k)}$ which we
shall denote, by a slight abuse of notation (since we are speaking 
about vertical parts with respect to the splitting \eqref{jet 
connection}), by 
$VJ_{s}\cA^{(r,k)}$. 

\bLm
Let $\chi(\lam,\mathfrak{G}(\bar{\Xi})_{V})\byd 
C^{1}_{1} (j_{2s}\hat{\Xi}\ten K_{hd\cL_{j_{2s}\bar{\Xi}_V}\lam})\equiv E_{j_{s}\hat{\Xi}\rfloor
hd\cL_{j_{2s+1}\bar{\Xi}_V}\lam}$. Let 
$D_{H}$ be the horizontal differential on $\bY_{\zet}\ucar{\bX}V\cA^{(r,k)}$. 
Then we have:
\beq
(\pi^{4s+1}_{2s+1})^{*}\chi(\lam,\mathfrak{G}(\bar{\Xi})_{V}) = E_{\chi(\lam,\mathfrak{G}(\bar{\Xi})_{V})} +
F_{\chi(\lam,\mathfrak{G}(\bar{\Xi})_{V})}\,,
\eeq 
where
\bEq
E_{\chi(\lam,\mathfrak{G}(\bar{\Xi})_{V}}: J_{4s}\bY_{\zet}\ucar{\bX}VJ_{4s}\cA^{(r,k)} \to 
\Con^{*}_{0}[\cA^{(r,k)}]\ten\Con^{*}_{0}[\cA^{(r,k)}]\wed (\owed{n}T^{*}\bX) \,,
\eEq
and locally, $F_{\chi(\lam,\mathfrak{G}(\bar{\Xi})_{V})} = 
D_{H}M_{\chi(\lam,\mathfrak{G}(\bar{\Xi})_{V})}$, with
\beq
M_{\chi(\lam,\mathfrak{G}(\bar{\Xi})_{V})}:J_{4s}\bY_{\zet}\ucar{\bX}VJ_{4s}\cA^{(r,k)} \to 
\Con^{*}_{2s-1}[\cA^{(r,k)}]\ten \Con^{*}_{0}[\cA^{(r,k)}]\wed
(\owed{n-1}T^{*}\bX)\,.
\eeq
\eLm

\bPf
Notice that, since $\chi(\lam,\mathfrak{G}(\bar{\Xi})_{V})\equiv
E_{(-\pounds_{j_{s}\Xi_{V}}\rfloor hd\cL_{j_{2s+1}\bar{\Xi}_V}\lam)}$, as a consequence of the
linearity  properties
of $\chi(\lam, \Xi)$ and 
of linearity properties of the Lie derivative operator $\pounds$ (see 
Subsections 
$2.3$ and $2.4$) we have $\chi(\lam,\mathfrak{G}(\bar{\Xi})_{V}): 
J_{2s}\bY_{\zet}\ucar{\bX}VJ_{2s}\cA^{(r,k)} \to
\Con^{*}_{2s}[\bY_{\zet}\ucar{\bX}V\cA^{(r,k)}]$ $\ten$ 
$\Con^{*}_{0}[\cA^{(r,k)}]$ $\wed$  $(\owed{n}T^{*}\bX)$ and 
$D_{H}\chi(\lam,\mathfrak{G}(\bar{\Xi})_{V})$ $=$ $0$. 
Thus the decomposition Lemma \ref{kolar} can be applied.\QED
\ePf

\bDf
Let $\bar{\Xi}\in \cA^{(r,k)}$. \\
We call the morphism $\cJ(\lam,\mathfrak{G}(\bar{\Xi})_{V})$ $\byd$
$E_{\chi(\lam,\mathfrak{G}(\bar{\Xi})_{V})}$  the {\em gauge-natural generalized Jacobi 
morphism} associated with the Lagrangian $\lam$ and the gauge-natural lift $\mathfrak{G}(\bar{\Xi})_{V}$.
\END\eDf

We have the following:
\bPr
The morphism $\cJ(\lam,\mathfrak{G}(\bar{\Xi})_{V})$ is a 
{\em linear} morphism with respect to the projection 
$J_{4s}\bY_{\zet}\ucar{\bX}VJ_{4s}\cA^{(r,k)} \to J_{4s}\bY_{\zet}$.
\ePr

We are now able to provide an important specialization of Theorem \ref{GeneralJacobi}.
\bPr\label{GeneralJacobi2} 
Let $[\cL_{j_{s+1}\bar{\Xi}_{V}}\lam]\in (\Var^{n+1}_{s})_{\bY}$. Then we have
\beq
\kap^{4s+1}_{s+1} \com \cL_{j_{s}\bar{\Xi}} [\cL_{j_{s+1}\bar{\Xi}_{V}}\lam] = 
\cE_{n}({j_{s}\bar{\Xi}_{V} \rfloor h(\cL_{j_{s+1}\bar{\Xi}_{V}}\lam)}) 
+ \cJ(\lam,\mathfrak{G}(\bar{\Xi})_{V})\,.
\eeq
\ePr

\bPf
By Theorem \ref{GeneralJacobi} and the Lemma above we have:
\bEq
& & \kap^{4s+1}_{s+1} \com \cL_{j_{s}\bar{\Xi}} [\cL_{j_{s+1}\bar{\Xi}_{V}}\lam]
= \cE_{n}({j_{s}\bar{\Xi}_{V} \rfloor h(\cL_{j_{s+1}\bar{\Xi}_{V}}\lam)}) + \\
& & +[C^{1}_{1} (j_{2s}\bar{\Xi}\ten K_{hd\cL_{j_{2s}\bar{\Xi}_V}\lam})] = 
\cE_{n}({j_{s}\bar{\Xi}_{V} \rfloor h(\cL_{j_{s+1}\bar{\Xi}_{V}}\lam)}) + \\
& &\cE_{n}({j_{s}\bar{\Xi}_{V} \rfloor 
h(d\cL_{j_{s+1}\bar{\Xi}_{V}}\lam)})=
\cE_{n}({j_{s}\bar{\Xi}_{V} \rfloor
h(\cL_{j_{s+1}\bar{\Xi}_{V}}\lam)}) + \\
& &+ \cJ(\lam,\mathfrak{G}(\bar{\Xi})_{V})\,.\QED
\eEq
\ePf

\bRm\label{remark}
Theorem \ref{noether I} in Subsection $2.6$ provides an {\em invariant} decomposition, 
where both pieces are globally defined. However, the second one is only {\em 
locally} a divergence, unless some further geometric structures such 
as linear symmetric connections on the basis manifold or 
suitable gauge-natural principal (or prolongations with respect to 
linear symmetric connections of principal) connections 
are introduced \cite{Kol83,Kol84,Vit98}. 
Proposition \ref{GeneralJacobi2} above, {\em instead}, provides an invariant decomposition 
into two pieces which are globally defined and no one of them, {\em seen as a section of
$(\Var^{n+1}_{s})_{\bY}$}, is a (local) divergence. As we shall see, this
fact has very important  consequences concerning conserved quantities in gauge-natural 
Lagrangian field theories. 
\END\eRm

A simple comparison of Remark \ref{boh}, Proposition \ref{x} and the 
Proposition above gives us the following.
\bCr\label{comparison}
Let $\del^{2}_{\mathfrak{G}}\lam$ be the variation of $\lam$ with respect to vertical parts
of gauge-natural lifts of infinitesimal principal automorphisms. We have:
\bEq
\mathfrak{G}(\bar{\Xi})_{V}\rfloor
\cE_{n}(\mathfrak{G}(\bar{\Xi})_{V}\rfloor\cE_{n}(\lam))
=
\del^{2}_{\mathfrak{G}}\lam 
=
\cE_{n}(\mathfrak{G}(\bar{\Xi})_{V}\rfloor 
h(d\del\lam))\,. 
\eEq
\eCr

The reader should notice that, {\em seen as a section of
$(\Var^{n}_{s})_{\bY\ucar{\bX}V\bY}$}, the equivalence class 
$[\cE_{n}({j_{s}\bar{\Xi}_{V}
\rfloor  h(\del\lam)})]$ vanishes being a 
local divergence of higher degree contact forms. This result can also be compared with
\cite{FPV02}.

\section{Noether Theorems and conserved currents for gauge-natural invariant 
Lagrangians}

In the following we assume that the field equations are generated by 
means of a variational principle from a Lagrangian which is 
gauge-natural invariant, \ie invariant with respect to any 
gauge-natural lift of infinitesimal right invariant vector fields.
We consider now a projectable vector field $(\hat{\Xi},\xi)$ on 
$\bY_{\zet}$ and take into account the Lie derivative with respect to 
its prolongation $j_{s}\hat{\Xi}$.

\bDf\label{gn}
Let $(\hat{\Xi},\xi)$ be a projectable vector field on $\bY_{\zet}$. 
Let $\lam \in \Var^{n}_{s}$ 
be a generalized Lagrangian. We say $\hat{\Xi}$ to be a {\em symmetry\/} 
of $\lam$ if $\cL_{j_{s+1}\hat{\Xi}}\,\lam = 0$.

We say $\lam$ to be a 
{\em gauge-natural invariant Lagrangian} if the gauge-natural lift 
$(\hat{\Xi},\xi)$ of {\em any} vector 
field $\bar{\Xi} \in \cA^{(r,k)}$ is a  symmetry for 
$\lam$, {\em i.e.} if $\cL_{j_{s+1}\bar{\Xi}}\,\lam = 0$. 
In this case the projectable vector field 
$\hat{\Xi}\equiv \mathfrak{G}(\bar{\Xi})$ is 
called a {\em gauge-natural symmetry} of $\lam$.\END 
\eDf

\bRm
Due to $\cE_{n}\cL_{j_{s}\bar{\Xi}} = \cL_{j_{s}\bar{\Xi}}\cE_{n}$, a symmetry of a Lagrangian
$\lam$ is also a symmetry of its Euler--Lagrange morphism $E_\lam$
(but the converse is not true, see {\em e.g.} \cite{Tra67}).\END
\eRm

Symmetries of a Lagrangian $\lam$ are calculated by means of 
Noether Theorems, which takes a particularly interesting 
form in the case of gauge-natural Lagrangians.  

\bPr
\label{symmetry of L}
Let $\lam \in \Var^{n}_{s}$ be a gauge-natural Lagrangian and 
$(\hat{\Xi},\xi)$ 
a gauge-natural symmetry of $\lam$. Then we have
\bEq\label{first decomposition}
0 = - \pounds_{\bar{\Xi}} \rfloor \cE_{n}(\lam) 
+d_{H}(-j_{s}\pounds_{\bar{\Xi}} 
\rfloor p_{d_{V}\lam}+ \xi \rfloor \lam) \,.
\eEq
Suppose that the section $\sig$ fulfills the 
condition
$(j_{2s+1}\sig)^{*}(- \pounds_{\bar{\Xi}} \rfloor \cE_{n}(\lam)) = 0$.
Then, the $(n-1)$--form 
\bEq
\eps = - j_{s}\pounds_{\bar{\Xi}} \rfloor p_{d_{V}\lam}+ \xi 
\rfloor \lam \label{current}\,,
\eEq
fulfills the equation $d ((j_{2s}\sig)^{*}(\eps)) = 0$.
\ePr

\bRm
If $\sig$ is a critical section for $\cE_{n}(\lam)$, \ie
$(j_{2s+1}\sig)^{*}\cE_{n}(\lam) = 0$, the above equation 
admits a physical interpretation as a so-called {\em weak conservation law} 
for the density associated with $\eps$.
\eRm

\bDf
Let $\lam \in \Var^{n}_{s}$ be a gauge-natural Lagrangian and 
$\bar{\Xi} \in \cA^{(r,k)}$. Then the sheaf morphism $\eps : 
J_{2s}\bY_{\zet} \ucar{\bX} VJ_{2s}\cA^{(r,k)}
\to \cC_{0}^{*}[\cA^{(r,k)}]\wed (\owed{n-1} T^{*}\bX)$
is said to be a {\em gauge-natural weakly conserved current\/}.\END 
\eDf

\bRm\label{arbitrary1}
In general, this conserved current is not uniquely defined. In fact, 
it depends on the choice of $p_{d_{V}\lam}$, which is not unique (see 
\cite{Vit98} and references quoted therein).
Moreover, we could add to the conserved current any form
$\mu \in \Var^{n-1}_{2s}$ which is variationally closed, \ie such 
that $\cE_{n-1}(\mu) = 0$ holds. The form $\mu$ is 
locally of the type $\mu = d_{H}\gam$, where $\gam \in 
\Var^{n-2}_{2s-1}$.
\END\eRm

\bCr\label{horizontal diff}
Let $\eps : J_{2s}\bY_{\zet} \ucar{\bX} VJ_{2s}\cA^{(r,k)}
\to \cC_{0}^{*}[\cA^{(r,k)}]\wed (\owed{n-1} T^{*}\bX)$ be a conserved current. 
As an immediate consequence of Remark \ref{Fund} we can regard $\eps$ as the 
equivalent morphism $\eps\equiv \overline{\eps}: 
J_{2s}\bY_{\zet} \ucar{\bX} VJ_{2s}\cA^{(r,k)} \to \cC_{2s}^{*}[\cA^{(r,k)}]
\ten\cC_{0}^{*}[\cA^{(r,k)}]
\wed (\owed{n-1} T^{*}\bX)$.
\eCr 

\bRm
Let $\eta \in \Var^{n+1}_{s}$ and let $\Xi$ be a symmetry of $\eta$.
Then, as a special case of Theorem \ref{GeneralJacobi}, we have
$0 = \cE_{n}(\Xi_{V} \rfloor \eta) +
[C^{1}_{1}(j_{2s+1}\bar{\Xi}_{V}\ten K_{hd\eta})]$.
Suppose that $K_{hd\eta} = 0$; then we have
$\cE_{n}(\bar{\Xi}_{V} \rfloor \eta) = 0$.
This implies that $\bar{\Xi}_{V} \rfloor \eta$ is variationally trivial, \ie 
it is locally of the type
$\bar{\Xi}_{V} \rfloor \eta = d_{H}\mu$, where $\mu \in \Var^{n-1}_{s-1}$.

Suppose that the section $\sig:\bX \to \bY_{\zet}$ fulfils
$(j_{2s+1}\sig)^{*}(\bar{\Xi}_{V} \rfloor \eta) = 0$. Then we have
$d((j_{2s}\sig)^{*} \mu) = 0$
so that, as in the case of symmetries of Lagrangians, 
$\mu$ is a conserved current along $\sig$.

As in the case of Lagrangians, a conserved current for an
Euler--Lagrange type morphism is not uniquely defined. In fact, we could add
to $\Xi_{V} \rfloor \eta$ any variationally trivial Lagrangian,
obtaining different conserved currents. Moreover, such conserved
currents are defined up to variationally trivial $(n-1)$--forms.\END
\eRm

\subsection{The Bianchi morphism}label{key1}

In gauge-natural Lagrangian theories it is a well known procedure  
to perform suitable integrations by 
parts to decompose the conserved current $\eps$ into the sum of 
a conserved current vanishing along solutions of the Euler--Lagrange equations, 
the so--called {\em reduced current}, and the
formal divergence of a skew--symmetric (tensor) density called a {\em 
superpotential} (which is defined modulo a divergence). 
Within such a procedure, the generalized Bianchi identities play a very fundamental role: they
are in fact necessary and (locally) sufficient conditions for the conserved current
$\epsilon$ to be not only closed but also the divergence of a skew-symmetric (tensor) density
along solutions of the  Euler--Lagrange equations.
  
In the following we shall perform such an integration by part of the conserved current by resorting to Kol\'a\v r's invariant
decomposition formula of vertical morphisms we already used to define the Jacobi
morphism. We will also make an extensive use of Remark \ref{Fund}.

\bRm
Let $\lam$ be a gauge-natural Lagrangian. By the linearity of
$\pounds$ we have 
\beq
\ome (\lam,\mathfrak{G}(\bar{\Xi})_{V}) = \pounds_{\bar{\Xi}} \rfloor \cE_{n} (\lam): J_{2s}\bY_{\zet}
\to \Con_{2s}^{*}[\cA^{(r,k)}]\ten \Con_{0}^{*}[\cA^{(r,k)}]\wed (\owed{n} 
T^{*}\bX) \,.
\eeq
We have $D_{H}\ome(\lam,\mathfrak{G}(\bar{\Xi})_{V})= 0$.
We can regard $\ome(\lam,\mathfrak{G}(\bar{\Xi})_{V})$ as the extended morphism 
$\ome(\lam,\mathfrak{G}(\bar{\Xi})_{V}): J_{2s}\bY_{\zet} \ucar{\bX} VJ_{2s}\cA^{(r,k)}
\to\Con_{2s}^{*}[\cA^{(r,k)}]\ten \Con_{2s}^{*}[\cA^{(r,k)}]\ten\Con_{0}^{*}[\cA^{(r,k)}]\wed
(\owed{n}T^{*}\bX)$. Thus we can state the following.\END\eRm

\bLm\label{kol}
Let $\ome(\lam,\mathfrak{G}(\bar{\Xi})_{V})$ be as in the above Remark.  Then we have {\em globally}
\beq
(\pi^{4s+1}_{s+1})^{*}\ome(\lam,\mathfrak{G}(\bar{\Xi})_{V}) = \bet(\lam,\mathfrak{G}(\bar{\Xi})_{V}) +
F_{\ome(\lam,\mathfrak{G}(\bar{\Xi})_{V})}\,,
\eeq 
where
\bEq\label{key}
& \bet(\lam,\mathfrak{G}(\bar{\Xi})_{V})
\equiv
E_{\ome(\lam,\mathfrak{G}(\bar{\Xi})_{V})} :\\
& :J_{4s}\bY_{\zet} \ucar{\bX} VJ_{4s}\cA^{(r,k)}
\to\Con_{2s}^{*}[\cA^{(r,k)}]\ten
\Con_{0}^{*}[\cA^{(r,k)}]\ten\Con_{0}^{*}[\cA^{(r,k)}]\wed
(\owed{n}T^{*}\bX)
\eEq
and {\em locally}, $F_{\ome(\lam,\mathfrak{G}(\bar{\Xi})_{V})} = D_{H}M_{\ome(\lam,\mathfrak{G}(\bar{\Xi})_{V})}$,
with 
\bEq
&M_{\ome(\lam,\mathfrak{G}(\bar{\Xi})_{V})}: J_{4s-1}\bY_{\zet} \ucar{\bX} VJ_{4s-1}\cA^{(r,k)} \to \\
& \to
\Con_{2s}^{*}[\cA^{(r,k)}]\ten
\Con_{2s-1}^{*}[\cA^{(r,k)}]\ten\Con_{0}^{*}[\cA^{(r,k)}]\wed 
(\owed{n-1}T^{*}\bX)\,.
\eEq
\eLm
In particular, we get the following {\em local} decomposition of $\ome(\lam,\mathfrak{G}(\bar{\Xi})_{V})$:
\bEq
\ome(\lam,\mathfrak{G}(\bar{\Xi})_{V}) = \bet(\lam,\mathfrak{G}(\bar{\Xi})_{V}) +
D_{H}\tilde{\eps}(\lam,\mathfrak{G}(\bar{\Xi})_{V}) \,,
\eEq

\bPf
We take into account that $D_{H}\ome(\lam,\mathfrak{G}(\bar{\Xi})_{V})$ is obviously
vanishing, then the  result is a straightforward consequence of Lemma \ref{kolar}.\QED
\ePf

\bDf
We call the global morphism $\bet(\lam,\mathfrak{G}(\bar{\Xi})_{V}) \byd E_{\ome(\lam,\mathfrak{G}(\bar{\Xi})_{V})}$ 
the {\em generalized} {\em Bianchi morphism} associated
with the Lagrangian $\lam$.
\END\eDf

\bRm
For any $(\bar{\Xi},\xi)\in \cA^{(r,k)}$, as a consequence of the gauge-natural invariance of the Lagrangian, 
the morphism $\bet(\lam,\mathfrak{G}(\bar{\Xi})_{V}) \equiv
\cE_{n}(\ome(\lam,\mathfrak{G}(\bar{\Xi})_{V}))$ is {\em locally} identically vanishing. 
We stress that these are just {\em local generalized Bianchi 
identities}.
In particular, we have $\ome(\lam,\mathfrak{G}(\bar{\Xi})_{V}) =
D_{H}\tilde{\eps}(\lam,\mathfrak{G}(\bar{\Xi})_{V})$ {\em locally} 
\cite{AnBe51,Ber49,Ber58,Gol58,Kom59}.\END
\eRm

\bDf 
The form $\tilde{\eps}(\lam,\mathfrak{G}(\bar{\Xi})_{V})$ is called a {\em local reduced current}.\END  
\eDf

\subsection{Global generalized Bianchi identities}

We are now able to state our main result providing necessary and sufficient 
conditions on the gauge-natural
lift of infinitesimal right-invariant automorphisms of the principal bundle 
$\bP$ in order to get globally defined generalized Bianchi
identities. 
Let $\mathfrak{K} \byd \textstyle{Ker}_{\cJ(\lam,\mathfrak{G}(\bar{\Xi})_{V})}$ 
be the {\em kernel} of
the generalized gauge-natural morphism $\cJ(\lam,\mathfrak{G}(\bar{\Xi})_{V})$. 
As a consequence of the considerations
above, we have the following important result.

\bTh
The generalized Bianchi morphism is globally vanishing if and only 
if $\del^{2}_{\mathfrak{G}}\lam\equiv\cJ(\lam,\mathfrak{G}(\bar{\Xi})_{V})= 
0$, \ie if and only if
$\mathfrak{G}(\bar{\Xi})_{V}\in\mathfrak{K}$.
\eTh

\bPf
By Corollary \ref{comparison} we get 
\beq
\mathfrak{G}(\bar{\Xi})_{V}\rfloor\bet(\lam,\mathfrak{G}(\bar{\Xi})_{V}) = 
\del^{2}_{\mathfrak{G}}\lam\equiv\cJ(\lam,\mathfrak{G}(\bar{\Xi})_{V})\,.
\eeq
Now, if $\mathfrak{G}(\bar{\Xi})_{V}\in\mathfrak{K}$ then
$\bet(\lam,\mathfrak{G}(\bar{\Xi})_{V})\equiv 0$, which are {\em global generalized Bianchi identities}. 
{\em Vice versa}, if $\mathfrak{G}(\bar{\Xi})_{V}$ is
such that $\bet(\lam,\mathfrak{G}(\bar{\Xi})_{V})= 0$, then
$\cJ(\lam,\mathfrak{G}(\bar{\Xi})_{V})$ $=$ $0$ and
$\mathfrak{G}(\bar{\Xi})_{V}\in\mathfrak{K}$. {\em Notice} that
$\mathfrak{G}(\bar{\Xi})_{V}\rfloor\bet(\lam,\mathfrak{G}(\bar{\Xi})_{V})$ is nothing but the
{\em Hessian morphism} associated with $\lam$ (see \cite{FPV02}).
\QED
\ePf

\bRm
We recall that given a vector field 
$j_{s}\hat{\Xi}: J_{s}\bY_{\zet} \to TJ_{s}\bY_{\zet}$, the splitting
\eqref{jet connection} yields $j_{s}\hat{\Xi} \, \com \, \pi^{s+1}_{s} = 
j_{s}\hat{\Xi}_{H} + j_{s}\hat{\Xi}_{V}$
where, if $j_{s}\hat{\Xi} = \hat{\Xi}^{\gam}\der_{\gam} + \hat{\Xi}^i_{\balp}\der^{\balp}_i$, then we
have $j_{s}\hat{\Xi}_{H} = \hat{\Xi}^{\gam}D_{\gam}$ and
$j_{s}\hat{\Xi}_{V} = 
D_{\balp}(\hat{\Xi}^i - y^i_{\gam}\hat{\Xi}^{\gam}) \der^{\balp}_{i}$. Analogous considerations
hold true of course also for the unique corresponding invariant vector 
field $j_{s}\bar{\Xi}$ on $W^{(r,k)}\bP$.
In particular, the condition $j_{s}\bar{\Xi}_{V}\in \mathfrak{K}$ implies, of course, that the components $\bar{\Xi}^i_{\balp}$
and
$\bar{\Xi}^{\gam}$ {\em are not} independent, but they are {\em related} in such a way that $j_{s}\bar{\Xi}_{V}$ must be a
solution of generalized gauge-natural Jacobi equations for the Lagrangian
$\lam$ (see coordinate expression \eqref{JJJ}). The geometric interpretation of this condition will be the subject of a
forthcoming paper
\cite{PaWi04}. Our results are quite
evidently related to the theory of
$G$-reductive Lie derivatives developed in \cite{GoMa03}. It is in fact our opinion that the Kosmann lift (the kind of
gauge-natural lift used to correctly define the Lie derivative of spinors, in \cite{GoMa03} interpreted as a special kind of
reductive lift) can be recognized as a kind of gauge-natural Jacobi vector field. Even more, we believe that the kernel
of the generalized gauge-natural Jacobi morphism induces a canonical reductive pair on $W^{(r,k)}\bP$. We also remark that,
for each $\bar{\Xi}\in \cA^{(r,k)}$ such that $\bar{\Xi}_{V}\in \mathfrak{K}$, we have
$\cL_{j_{s}\bar{\Xi}_{H}}\ome(\lam,
\mathfrak{K})=0$. Here it is enough to stress that, {\em within} a gauge-natural {\em
invariant Lagrangian variational principle} the gauge-natural lift of  infinitesimal
principal automorphism {\em is not} intrinsically arbitrary. It would be also interesting to
compare such results with reduction theorems stated in \cite{Ja03}.\END
\eRm 

\medskip

In the following we shall refer to {\em canonical} globally defined objects (such as currents or corresponding
superpotentials) by their explicit dependence on $\mathfrak{K}$.

\bCr
Let $\lam \in \Var^{n}_{s}$ be a gauge-natural Lagrangian and 
$j_{s}\hat{\Xi}_{V}\in \mathfrak{K}$ 
a gauge-natural symmetry of $\lam$. 
Being $\bet(\lam, \mathfrak{K})\equiv 0$, we have, {\em globally}, $\ome(\lam,\mathfrak{K}) = D_{H}\eps(\lam, \mathfrak{K})$,
then the following holds:
\bEq\label{strong conservation}
D_{H}(\eps(\lam, \mathfrak{K})-\tilde{\eps}(\lam, \mathfrak{K}) = 0\,.
\eEq
\eCr

Eq. \eqref{strong conservation} is referred as a gauge-natural 
`strong' conservation law for the {\em global} density $\eps(\lam, \mathfrak{K}) -\tilde{\eps}(\lam,
\mathfrak{K})$.

We can now state the following fundamental result about the existence and 
{\em globality} of gauge-natural superpotentials in the framework 
of variational sequences.

\bTh\label{global}
Let $\lam \in \Var^{n}_{s}$ be a gauge-natural Lagrangian and 
$(j_{s}\hat{\Xi},\xi)$ a gauge-natural symmetry of $\lam$. Then there exists 
a global sheaf morphism 
$\nu(\lam, \mathfrak{K})$ $\in$ $\left(\Var^{n-2}_{2s-1}\right)_{\bY_{\zet} \ucar{\bX} 
\mathfrak{K}}$
such that
\beq
D_{H}\nu(\lam, \mathfrak{K}) = \eps(\lam, \mathfrak{K}) -\tilde{\eps}(\lam,
\mathfrak{K})\,.
\eeq
\eTh

\bDf
We define the sheaf morphism $\nu(\lam, \mathfrak{K})$ to be a 
{\em canonical gauge-natural 
superpotential} associated with $\lam$.
\END\eDf

{\em Acknowledgments.}
The authors wish to thank Prof. I. Kol\'a\v r for 
many interesting and useful discussions. 


\end{document}